\documentclass[journal]{IEEEtran}
\usepackage{amsmath}
\usepackage{latexsym}
\usepackage[ruled,linesnumbered,noresetcount,vlined]{algorithm2e}

\SetKwFor{Repeat}{repeat}{:}{endw}

\usepackage{mdframed}
\usepackage{xspace}
\usepackage[export]{adjustbox}

\usepackage{authblk}
\usepackage{float}
\floatstyle{ruled}
\newfloat{model}{thp}{lop}
\floatname{model}{Model}

\usepackage{multicol}

\usepackage{amsmath,amsfonts,mathrsfs,amssymb,bm,mathtools,mathabx}
\usepackage{times, helvet, courier}
\usepackage{graphicx, epsfig, epstopdf}
\usepackage{wrapfig, multirow, multicol}
\usepackage{rotating, booktabs,longtable}
\usepackage{url}
\usepackage{stmaryrd,xspace,arydshln}
\usepackage{picinpar}
\usepackage{todonotes}
\usepackage{color}
\usepackage{hyperref}

\usepackage{xcolor,colortbl}
\usepackage{subcaption,cleveref}

\def\thmspace{0.2em}

\newcommand{\benumerate}{\begin{list}{$\bullet$}{\topsep=0pt \parsep=0pt \itemsep=1pt \labelwidth=1.5em \labelsep=0.5em \leftmargin=20pt}}
\newcommand{\eenumerate}{\end{list}}
\newcommand{\bitemize}{\begin{list}{$\bullet$}{\topsep=0pt \parsep=0pt \itemsep=1pt \leftmargin=10pt}}
\newcommand{\eitemize}{\end{list}}

\def\lvec{\bm{\left(}}
\def\rvec{\bm{\right)}}
\newcommand{\vect}[1]{{\lvec{#1}\rvec}}

\let\oldnl\nl
\newcommand{\nonl}{\renewcommand{\nl}{\let\nl\oldnl}}

\newcommand{\argmin}{\operatornamewithlimits{argmin}}

\newcommand{\setf}[1]{{\bf{#1}}}

\newcommand{\cO}{{\mathcal{O}}}	%

\newcommand{\bx}{\setf{x}}
\newcommand{\by}{\setf{y}}
\newcommand{\bw}{\setf{w}}

\newcommand{\blambda}{\bm{\lambda}}

\newcommand{\RR}{\mathbb{R}}

\pdfoutput=1
\usepackage{graphicx}
\graphicspath{{images/}}
\usepackage[section]{placeins}
\usepackage{enumitem,kantlipsum}

\begin{document}
\title{High-Fidelity Machine Learning Approximations \\ of Large-Scale Optimal Power Flow}

\author{Minas Chatzos, Ferdinando~Fioretto,
        Terrence~W.K.~Mak,~\IEEEmembership{Member,~IEEE,} 
        and~Pascal~Van~Hentenryck,~\IEEEmembership{Member,~IEEE}
\thanks{
Chatzos, Mak and Van Hentenryck are affiliated with the 
School of Industrial and Systems Engineering,
Georgia Institute of Technology, Atlanta, GA 30332, USA. Fioretto is affiliated with the Electrical Engineering And Computer Science Department, Syracuse University, Syracuse, NY 13244, USA. E-mail: minas@gatech.edu, ffiorett@syr.edu, wmak@gatech.edu, pvh@isye.gatech.edu.}
}

\markboth{}%
{}

\maketitle\sloppy\allowdisplaybreaks

\begin{abstract}
The AC Optimal Power Flow (AC-OPF) is a key building block in many
power system applications. It determines generator setpoints at
minimal cost that meet the power demands while satisfying the
underlying physical and operational constraints. It is non-convex and
NP-hard, and computationally challenging for large-scale power
systems.  Motivated by the increased stochasticity in generation
schedules and increasing penetration of renewable sources, this paper
explores a deep learning approach to deliver highly efficient and
accurate approximations to the AC-OPF. In particular, the paper
proposes an integration of deep neural networks and Lagrangian duality
to capture the physical and operational constraints.  The resulting
model, called OPF-DNN, is evaluated on real case studies from the
French transmission system, with up to 3,400 buses and 4,500
lines. Computational results show that OPF-DNN produces, in
milliseconds, highly accurate AC-OPF approximations whose costs are
within 0.01\% of optimality and which capture the problem constraints
with high fidelity.
\end{abstract}


\IEEEpeerreviewmaketitle

\section{Introduction}

The \emph{AC Optimal Power Flow} (AC-OPF) determines the most
economical generation dispatch that meets the demands while satisfying
physical and engineering constraints \cite{OPF}. It is a non-convex
and NP-hard \cite{7063278} problem, and the building bock of many
applications, including security-constrained OPFs (Monticelli et
al.~\cite{monticelli:87}), optimal transmission switching \cite{OTS},
capacitor placement \cite{baran:89}, expansion planning (Verma et
al.~\cite{verma:16}), and security-constrained unit commitment
\cite{wang:08}.

This paper explores a Machine-Learning (ML) approach to the AC-OPF.
It is motivated by the recognition that the integration of renewable
energy in sub-transmission and distribution systems has introduced
significant stochasticity in front and behind the meter, making load
profiles much harder to predict and introducing significant variations
in load and generation. This uncertainty forces system operators to
adjust the generators' setpoints with increasing frequency. However,
the resolution frequency to solve OPFs is limited by the AC-OPF
runtimes.  As a result, system operators typically solve OPF
approximations such as the linearized DC model (DC-OPF).  While these
approximations are more efficient computationally, their solutions may
be sub-optimal and induce substantial financial losses
\cite{mak2018phase} or they may fail to satisfy the physical and
engineering constraints.  An ML approach to the AC-OPF is also
promising for expansion planning and other design problems, where
plans are typically evaluated by solving a massive number of
multi-year Monte-Carlo simulations at 15-minute intervals
\cite{pachenew,Highway50}. For computational reasons, modern
approaches recur to the linear DC-OPF approximation and focus only on
the scenarios considered most pertinent \cite{pachenew} at the expense
of the fidelity of the simulations.  Adopting ML to approximate AC-OPF
solutions may fundamentally increase the accuracy of these simulations
by handling the non-convexities and a larger collection of scenarios.

The ML approach explored in this paper approximates the AC-OPF using a
Deep Neural Network (DNN) model.  This approximation can be seen as an
empirical risk minimization problem. However, the resulting setpoints
must also satisfy the physical and engineering constraints that
regulate power flows, and these constraints introduce significant
difficulties for ML-based approaches, as shown in
\cite{ng2018statistical,deka:2019}. To address this challenge, this
paper presents a DNN approach (OPF-DNN) to the AC-OPF that borrows
ideas from Lagrangian duality. OPF-DNN models the learning task as the
Lagrangian dual of the empirical risk minimization problem under the
AC-OPF constraints.

The paper makes the following contributions. At the methodological
level, it shows how to model the physical and engineering constraints
of the AC-OPF during training by combining Lagrangian duality and a
DNN architecture.  This integration is a key element to produce AC-OPF
approximations of high fidelity, that satisfy more than 99\% of the
constraints and have small violations for the remaining ones.  At the
conceptual level, this paper shows, for the first time, that DNN
architectures can provide high-fidelity approximations for OPF
problems with thousands of buses and transmission lines, significantly
expanding prior results.  The proposed framework is evaluated on real
case studies based on the French transmission system and is shown to
produce highly accurate AC-OPF approximations in a few milliseconds.
The paper is a revised and extended version of our conference paper
\cite{Fioretto:dnnopf} with significant improvements in the algorithm
design, training procedures, and the application of OPF-DNN to
real-life case studies that are one order of magnitude larger than
those reported in the prior study.

\section{Related Work}


Within the energy research landscape, DNN architectures have mainly
been adopted to predict exogenous factors affecting renewable
resources, such as solar or wind.  For instance, Anwar et
al.~\cite{Anwar:16} uses a DNN-based system to predict wind speed and
adopt the predictions to schedule generation units ahead of the
trading period, and Boukelia et al.~\cite{Boukelia:17} studied a DDN
framework to predict the electricity costs of solar power plants
coupled with a fuel backup system and energy
storage. Chatziagorakis~\cite{Chatziagorakis:16} studied the control
of hybrid renewable energy systems, using recurrent neural networks to
forecast weather conditions.  Another power system area in which DNNs
have been adopted is that of \emph{security assessment}:
Ince~\cite{Ince:16} proposed a convolutional neural network (CNN)
model for real-time power system fault classification to detect
faulted power system voltage signals. Arteaga~\cite{Arteaga:19}
proposed a convolutional neural network to identify safe vs.~unsafe
operating points to reduce the risks of a
blackout. Donnot~\cite{donnot:hal-02268886} use a ResNet architecture
to predict the effect of interventions that reconnect disconnected
transmission lines.

The literature on predicting the AC-OPF is much sparser. A survey by
Hasan et al.~\cite{hasan20survey} summarize recent developments in the
area.  Pan et al.~\cite{pan19deepopf} explore DNN architectures for
predicting DC-OPFs.  Deka et al.~\cite{deka2019learning} and Ng et
al.~\cite{ng2018statistical} use a DNN architecture to learn the set
of active constraints.  By exploiting the linearity of the DC-OPF
problem, once the set of relevant active constraints is identified, an
exhaustive search can be used to find a solution that satisfies the
active constraints.  While this strategy is efficient when the number
of active constraints is small, it becomes computationally challenging
when the number of active constraints increases. Additionally, this
strategy only applies to the DC-OPF.  The deep learning approach in
\cite{8887248} predicts voltages and flows for the AC-OPF.  However,
this approach focuses on specific operational constraints but ignores
other physical and engineering constraints, and is only validated on
test cases with up to 600 buses.  Zamzam and
Baker~\cite{DBLP:journals/corr/abs-1910-01213} propose another
approach, which does not consider problem constraints and is validated
only on small test cases.  The OPF-DNN approach proposed in this paper
is validated on large and realistic transmission grids from the French
transmission system, and attains higher prediction quality compared to
existing approaches.  The approach predicts all the variables and
models all physical and engineering constraints through Lagrangian
duality, and provides validation results for case studies up to 3400
buses and 4500 lines.

\section{Preliminaries}

\emph{Variables} are in lowercase, \emph{constants} are denoted by
dotted symbols and \emph{vectors} by bold symbols. The hat notation
$\hat{x}$ describes the prediction of a value $x$ and $\|\cdot\|$ is
used to denote the $L_1$-norm.  The power flow equations are expressed
in terms of complex \emph{power} of the form $S \!=\! (p \!+\! jq)$,
where $p$ and $q$ denote active and reactive power, \emph{admittances}
of the form $Y \!=\! (g \!+\! jb)$, where $g$ and $b$ denote the
conductance and susceptance, and \emph{voltages} of the form $V \!=\!
(v \angle \theta)$, with magnitude $v$ and phase angle $\theta$.

\begin{model}[!t]
	{\footnotesize
	\caption{AC Optimal Power Flow (AC-OPF)}
	\label{model:ac_opf}
	\vspace{-6pt}
	\begin{flalign}
		&{\cO}(\dot{\bm{p}}^d, \dot{\bm{q}}^d) = 
		\textstyle \argmin_{\bm{p}^g,\bm{v}}\;\;
		 \sum_{i \in {\cal N}} 	\text{cost}(p^g_i) && \label{ac_obj} \\
		&\mbox{\bf subject to:} \notag\\
		&\hspace{6pt}
		\dot{v}^{\min}_i \leq v_i \leq \dot{v}^{\max}_i 		
		&& \!\!\!\!\!\forall i \in {\cal N} 		\label{con:2a} \tag{2a}\\
		&\hspace{6pt}
		\text{ -- }\dot{\theta}^\Delta_{ij} \leq \theta_i \text{ -- } \theta_j  \leq \dot{\theta}^\Delta_{ij} 	
		&& \!\!\!\!\!\forall (ij) \in {\cal E}  	 \label{con:2b}\!\!\!\!\! \tag{$2b$}\\
		&\hspace{6pt}
		\dot{p}^{g\min}_i \leq p^g_i \leq \dot{p}^{g\max}_i 	
		&& \!\!\!\!\!\forall i \in {\cal N} 		\label{con:3a} \tag{$3a$}\\
		&\hspace{6pt}
		\dot{q}^{g\min}_i \leq q^g_i \leq \dot{q}^{g\max}_i 	
		&& \!\!\!\!\!\forall i \in {\cal N} 		\label{con:3b} \tag{3b}\\
		&\hspace{6pt}
		(p_{ij}^f)^2 + (q_{ij}^f)^2 \leq \dot{S}^{f\max}_{ij}			
		&& \!\!\!\!\!\forall (ij) \in {\cal E}	\label{con:4}  \tag{$4$}\\
		&\hspace{6pt}
		p_{ij}^f \!=\! \dot{g}_{ij} v_i^2 \text{--}  
		v_i v_j (\dot{b}_{ij} \!\sin(\theta_i \text{--} \theta_j)
		+ \dot{g}_{ij} \!\cos(\theta_i \text{--} \theta_j)\!)	
		&& \!\!\!\!\!\forall (ij)\!\in\! {\cal E} 	\label{con:5a} \tag{$5a$}\\
		&\hspace{6pt} 
		q_{ij}^f \!=\! \text{--} \dot{b}_{ij} v_i^2 \text{--}  v_i v_j (\dot{g}_{ij} \!\sin(\theta_i \text{--} \theta_j)
		\text{--} \dot{b}_{ij} \!\cos(\theta_i \text{--} \theta_j)\!)	
		&& \!\!\!\!\!\forall (ij)\!\in\! {\cal E}		\label{con:5b} \tag{5b}\\
		&\hspace{6pt}
		p^g_i \text{ -- } \dot{p}^d_i = \textstyle \sum_{(ij)\in {\cal E}} p_{ij}^f	
		&& \!\!\!\!\!\forall i\in {\cal N} 		\label{con:6a} \tag{$6a$}\\
		&\hspace{6pt}
		q^g_i \text{ -- } \dot{q}^d_i = \textstyle 	\sum_{(ij)\in {\cal E}} q_{ij}^f	
		&& \!\!\!\!\!\forall i\in {\cal N} 		\label{con:6b} \tag{6b}\\
	&\textbf{output:}~~(\bm{p}^g, \bm{v}) \text{ -- The system operational parameters}
	\!\!\!\!\!\!\!\!\!\!\!\!\!\!\!\!\!\!\!\!
	\notag
	\end{flalign}
	\vspace{-12pt}
	}
\end{model}
\setcounter{equation}{6}

\paragraph{Optimal Power Flow}

The \emph{Optimal Power Flow (OPF)} determines the least-cost
generator dispatch that meets the load (demand) in a power network. A
power network is viewed as a graph $({\cal N}, {\cal E})$ where the
nodes $\cal N$ represent the set of $n$ \emph{buses} and the edges
$\cal E$ the set of $e$ \emph{transmission lines}. The OPF constraints
include physical and engineering constraints, which are captured in
the AC-OPF formulation of Model~\ref{model:ac_opf}.  The model uses
$\bm{p}^g$, and $\bm{p}^d$ to denote, respectively, the vectors of
active power generation and load associated with each bus and
$\bm{p}^f$ to describe the vector of active power flows associated
with each transmission line. Similar notations are used to denote the
vectors of reactive power $\bm{q}$.  Finally, the model uses $\bm{v}$
and $\bm{\theta}$ to describe the vectors of voltage magnitudes and
angles associated with each bus. The OPF takes as inputs the loads
$(\dot{\bm{p}}^d\!, \dot{\bm{q}}^d)$ and the admittance matrix
$\dot{\bm{Y}}$, with entries $\dot{g}_{ij}$ and $\dot{b}_{ij}$ for
each line $(ij) \!\in\!  {\cal E}$; It returns the active power vector
$\bm{p}$ of the generators, as well the voltage magnitude $\bm{v}$ at
the generator buses. The objective function \eqref{ac_obj} captures
the cost of the generator dispatch, and is typically expressed as a
linear or quadratic function. Constraints \eqref{con:2a} and
\eqref{con:2b} restrict the voltage magnitude and the phase angle
differences within their bounds.  Constraints \eqref{con:3a} and
\eqref{con:3b} enforce the generator active and reactive output
limits.  Constraints \eqref{con:4} enforce the line flow limits.
Constraints \eqref{con:5a} and \eqref{con:5b} capture \emph{Ohm's
  Law}. Finally, constraint \eqref{con:6a} and \eqref{con:6b} capture
\emph{Kirchhoff's Current Law} enforcing flow conservation.

\begin{model}[!t]
	{\small
	\caption{The Load Flow Model}
	\label{model:load_flow}
	\vspace{-6pt}
	\begin{flalign}
		\mbox{\bf minimize:}& \;\;
		\| \bm{p}^g - \hat{\bm{p}}^g \|^2 + \| \bm{v} - \hat{\bm{v}} \|^2 \label{load_flow_obj} \\
		\mbox{\bf subject to:} & \;\; \eqref{con:2a}-\eqref{con:6b} \notag
	\end{flalign}
	}
	\vspace{-12pt}
\end{model}

\paragraph{Deep Learning Models}

Supervised learning approximates a complex non-linear mapping from
labeled data.  Deep neural networks (DNNs) are ML models composed of a
sequence of layers, each typically taking as inputs the results of the
previous layer \cite{lecun2015deep}. Feed-forward neural networks are
basic DNNs where the layers are fully connected and the function
connecting the layers is given by
\[
\bm{o} = \pi(\bm{W} \bm{x} + \bm{b}),
\]
where $\bx \!\in\! \RR^n$ is the input vector, $\bm{o} \!\in\!  \RR^m$
the output vector, $\bm{W} \!\in\! \RR^{m \times n}$ a weight matrix,
and $\bm{b} \!\in\! \RR^m$ a bias vector. The function $\pi(\cdot)$ is
non-linear (e.g., a rectified linear unit (ReLU)).

The AC-OPF is an ideal candidate for supervised learning for a number of reasons:
\begin{enumerate}[wide, labelwidth=!,labelindent=0pt]
\item \textbf{Data availability:} A large amount of historical training
  data is available in industry.

\item\textbf{Input pattern:} The problem inputs are reasonably
  consistent, lie in specific intervals, and present spatial and
  temporal correlations between individual components. These
  properties make it possible to learn AC-OPFs without the need of a
  dataset that is exponential w.r.t.~the input size.
  
\item \textbf{Speed:} Once trained, a DNN can approximate the AC-OPF
  extremely fast, as only matrix operations are needed and GPU
  technology can be used. This opens a new realm of applications for
  AC-OPF that were hitherto limited because of computational challenges.

\item \textbf{Difficult instances:} In congested instances, the AC-OPF
  or its approximations (e.g., the DC model) may need significantly
  more time to produce a solution. This is not the case for a
  DNN-based model. Moreover, it has been shown experimentally that
  AC-OPF approximations result to large optimality gaps in real
  transmission systems \cite{mak2018phase}. These gaps do not occur
  using a DNN approach, as shown in Section \ref{sec:experiments}.
\end{enumerate}

\section{OPF Learning Goals}

Given loads $\vect{\bm{p}^d, \bm{q}^d}$, the goal is to predict the
control setpoints $\vect{\bm{p}^g, \bm{v}}$ of the generators.  The
resulting predictor, called OPF-DNN, learns an OPF mapping ${\cal O}:
\RR^{2n} \to \RR^{2m}$, where $n$ is the number of loads and $m$ is
the number of generators.  The input of the learning task is a dataset
${\cal D} \!=\! (\mathcal{X}, \mathcal{Y}) \!=\!
\{(\bx_\ell,\by_\ell)\}_{\ell\!=\!1}^N$, where $\bx_\ell \!\!=\!\!
(\bm{p}^d, \bm{q}^d)$ and $\by_\ell \!=\!  (\bm{p}^g, \bm{v})$
represents the $\ell^{th}$ observation of load demands and generator
setpoints that satisfy $\by_\ell \!=\! {\cal O}(\bx_\ell)$. The output
is a function $\hat{\cal O}$ that ideally would be the result of the
optimization problem
\begin{flalign*}
\mbox{\bf minimize:} & \;\; \sum_{\ell=1}^N {\cal L}_o(\by_\ell,\hat{\cal O}(\bx_\ell)) \\ 
\mbox{\bf subject to:} & \;\; {\cal C}(\bx_\ell,\hat{\cal O}(\bx_\ell))
\end{flalign*}
\noindent
where the loss function is specified by
\begin{equation}
\label{basic_loss}
	{\cal L}_o(\by, \hat{\by}) = 
	\underbrace{\| \bm{p}^g - \hat{\bm{p}}^g \|}_{{\cal L}_p(\by, \hat{\by})}{\!} +
	\underbrace{\| \bm{v} - \hat{\bm{v}} \|}_{{\cal L}_v(\by, \hat{\by})}{\!}
\end{equation}
and ${\cal C}(\bx,\by)$ holds if there exist voltage angles
$\bm{\theta}$ and reactive power $\bm{q}^g$ that produce a
feasible solution to the OPF constraints with $\bx =
(\bm{p}^d, \bm{q}^d)$ and $\by = (\bm{p}^g, \bm{v})$.

A baseline DNN model can obtained by ignoring the problem constraints
and minimizing the loss function.  It will produce an approximation
$\hat{\cal O}$ will typically not satisfy the OPF constraints. A key
challenge of the learning task is thus to design a DNN that
incorporate these constraints. 


\section{Modeling the OPF Constraints}

To capture OPF constraints, this paper uses a Lagrangian relaxation
approach based on constraint violations \cite{Fontaine:14} used in
generalized augmented Lagrangian relaxation \cite{Hestenes:69}. The
Lagrangian relaxation of an optimization problem
\begin{flalign*}
\mbox{\bf minimize:} & \;\; f(\bx) \\
\mbox{\bf subject to:} & \;\; h(\bx) = 0\\ 
                       &\;\; g(\bx) \leq 0 
\end{flalign*}

\noindent
is given by $\mbox{\bf minimize} f(\bx) + \lambda_h h(\bx) + \lambda_g g(\bx)$ 
where $\lambda_h$ and $\lambda_g \geq 0$ are the Lagrangian multipliers.
In contrast, the violation-based Lagrangian relaxation is 
\begin{flalign*}
\mbox{\bf minimize:} & \;\; f(\bx) + \lambda_h |h(\bx)| + \lambda_g \max(0,g(\bx))
\end{flalign*}
\noindent
with $\lambda_h,\lambda_g \geq 0$. In other words, the traditional
Lagrangian relaxation exploits the satisfiability degrees of
constraints, whereas the violation-based Lagrangian relaxation is
expressed in terms of violation degrees.  The satisfiability degree of
a constraint measures how well the constraint is satisfied, with
negative values representing the slack and positive values
representing violations. The violation degree is always non-negative
and represents how much the constraint is violated.  More formally,
the satisfiability degree of a constraint $c \!:\!  \mathbb{R}^n \to
\text{Bool}$ is a function $\sigma_c\!:\!  \mathbb{R}^n \to
\mathbb{R}$ such that $c(\bx)$ holds whenever $\sigma_c(\bx) \leq
0$. The violation degree of a constraint $c$ is a function $\nu_c\!:\!
\mathbb{R}^n \to \mathbb{R^+}$ such that $c(\bx)$ holds whenever
$\nu_c(\bx) \equiv 0$. For instance, for a linear constraints $c(\bx)$
of type $A\bx \geq b$, the \emph{satisfiability degree} is defined as
\begin{equation*}
\sigma_c(\bx) \equiv \bm{b} - A\bx
\end{equation*}
and the \emph{violation degrees} for inequality and equality 
constraints are specified by
\begin{equation*}
\nu^{\geq}_c(\bx) = \max\left(0, \sigma_c(\bx)\right) 
\qquad
\nu^{=}_c(\bx) = \left| \sigma_c(\bx) \right|. 
\end{equation*}
\noindent
Although the resulting term is not differentiable (but admits
subgradients), computational experiments indicated that violation
degrees are more appropriate for predicting OPFs than satisfiability
degrees. Observe also that an augmented Lagrangian method uses both
the satisfiability and violation degrees in its objective.

To define the violation degrees of the AC-OPF constraints, the
baseline model needs to be extended to predict the reactive power
dispatched $\bm{q}^g$ and the voltage angles $\bm{\theta}$.  Given the
predicted values $\hat{\bm{v}}, \hat{\bm{\theta}}, \hat{\bm{p}}^g,$
and $\hat{\bm{q}}^g$, the constraints can be captured naturally in
terms of satisfiability and violation degrees. For instance, the
satisfiability degree of a constraint in \eqref{con:4} can be
expressed as
\[
\sigma_4(\tilde{p}^f_{ij}, \tilde{q}^f_{ij}) = (\tilde{p}^f_{ij})^2 + (\tilde{q}^f_{ij})^2 - \dot{S}^{\max}_{ij}
\]
for all $(ij) \in {\cal E}$ and the violation degree becomes
\[
\nu_{4}(\tilde{\bm{p}}^f\!, \tilde{\bm{q}}^f) = \frac{1}{e} \sum_{(ij) \in {\cal E}} \nu_c^{\geq}\big(\sigma_{4}(\tilde{p}_{ij}^f, \tilde{q}_{ij}^f)\big).
\]

\section{Objective}

The loss function used to train OPF-DNN can now be derived
systematically.  First, the loss is augmented to consider the
predictions of voltage phase angles and the reactive power of
generators, since these are required to compute the violation degrees
associated with the OPF constraints. The resulting loss function
${\cal L}_o(\by, \hat{\by})$ is:
\begin{flalign}
	\label{obj_advanced}
	\underbrace{\| \bm{v} - \hat{\bm{v}} \|}_{{\cal L}_v(\by, \hat{\by})}{\!} + 
	\underbrace{\| \bm{\theta} - \hat{\bm{\theta}} \|}_{{\cal L}_\theta(\by, \hat{\by})}{\!} +
	\underbrace{\| \bm{p}^g - \hat{\bm{p}}^g \|}_{{\cal L}_p(\by, \hat{\by})}{\!} +
	\underbrace{\| \bm{q}^g - \hat{\bm{q}}^g \|}_{{\cal L}_q(\by, \hat{\by})}{\!}.
\end{flalign}
\noindent
It minimizes the mean $L_1$-error between the optimal voltage and
generator settings $\bm{y}$ and their predictions $\hat{\bm{y}}$.

\begin{figure*}[!tbh]
\centering\includegraphics[width=0.7\linewidth]{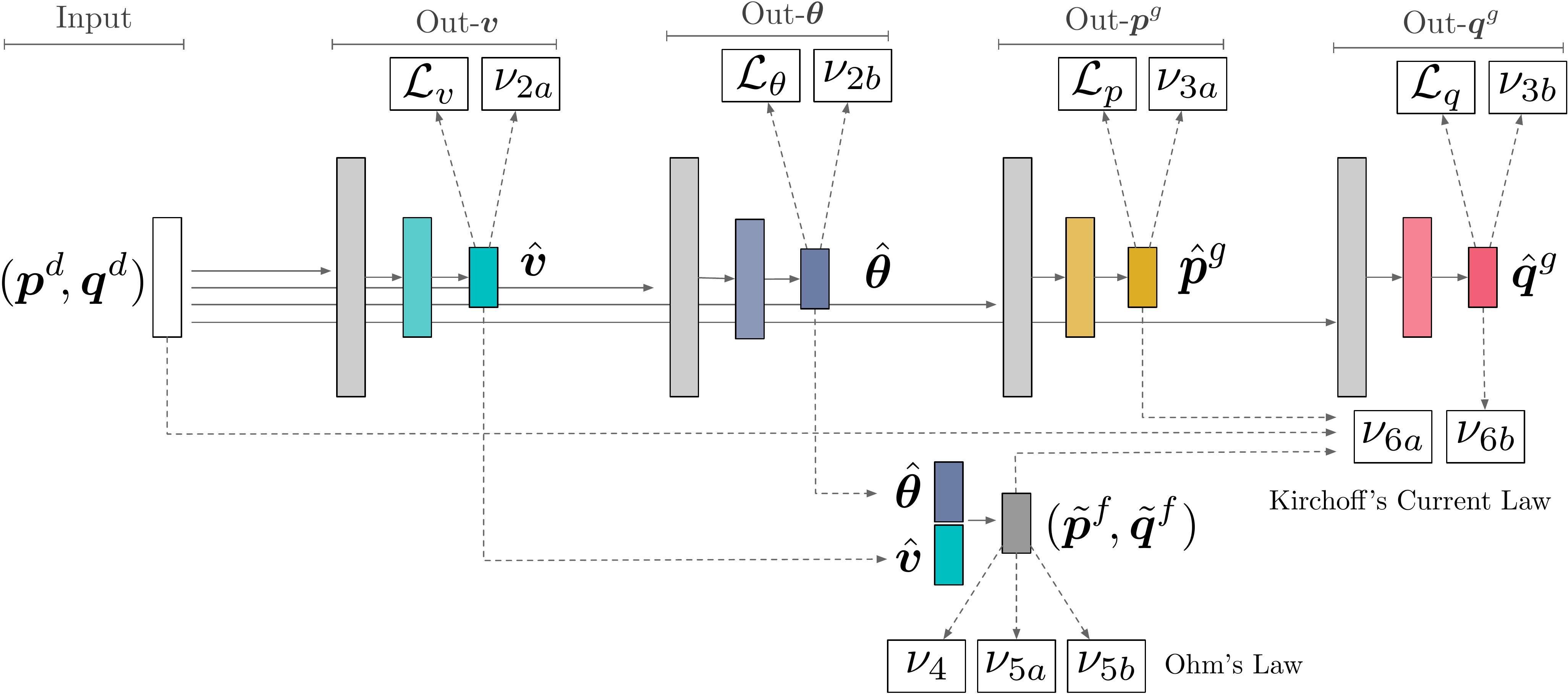}
\caption{The OPF-DNN Model:
		Each layer is fully connected with ReLU activation. 
		White boxes correspond to input tensors, dark-colored boxes 
		correspond to output layers. Loss components and violation 
		degrees are shown as white rectangles.}
\label{fig:dlopf} 
\end{figure*}

Moreover, the objective function includes the Lagrangian relaxation
that uses the violation degrees of the physical and engineering
constraints. Given the set ${\cal C}$ of OPF constraints, the
associated loss is captured by the expression
\begin{flalign*}
{\cal L}_c(\bx, \hat{\by}) &= \sum_{c \in {\cal C}} \lambda_c \nu_c(\bx,\hat{\by}),
\end{flalign*}
where $\nu_c(\bx,\hat{\by})$ is the violation degree of constraint $c$
for input $\bx$ and prediction $\hat{\by}$.  The model loss function
sums these two terms, i.e.,
\[
{\cal L}(\bx,\by,\hat{\by}) = {\cal L}_o(\by, \hat{\by}) + {\cal L}_c(\bx, \hat{\by}).
\]

\section{The Network Architecture}

The OPF-DNN architecture is outlined in Figure \ref{fig:dlopf}. Its
inputs are load vectors $(\bm{p}^d, \bm{q}^d)$. The network has four
basic units composed of fully connected layers with ReLU
activations. Each subnetwork predicts a target variable: voltage
magnitude $\hat{\bm{v}}$, phase angle $\hat{\bm{\theta}}$, active
power generation $\hat{\bm{p}}^g$, and reactive power generation
$\hat{\bm{q}}^g$. The predictions for the voltage magnitude
$\hat{\bm{v}}$ and angle $\hat{\bm{\theta}}$ are used to compute the
load flows $(\tilde{\bm{p}}^f, \tilde{\bm{q}}^f\!)$, as illustrated in
the bottom of Figure \ref{fig:dlopf}. Predicting the values
$(\tilde{\bm{p}}^f, \tilde{\bm{q}}^f\!)$ indirectly is beneficial for
two reasons. First, a direct prediction would require additional
components in the neural network topology, greatly increasing its
complexity, its training time, and memory utilization. These are of
critical importance for scaling to large-scale power systems. Second,
experimental results indicated that indirect flow predictions may
increase the overall accuracy of the predictions.

\section{Lagrangian Duality}

Let $\hat{\cal O}[\bw]$ be the resulting OPF-DNN with weights $\bw$
and let ${\cal L}[\blambda]$ be the loss function parametrized by the
Lagrangian multipliers $\blambda = \{\lambda_c\}_{c \in {\cal C}}$.
The training aims at finding the weights $\bw$ that minimize the loss
function for a given set of Lagrangian multipliers, i.e., it computes
\[
\text{LR}(\blambda) = \min_{\bw} {\cal L}[\blambda](\bx,\by,\hat{\cal O}[\bw](\bx)).
\]
It remains to determine appropriate Lagrangian multipliers, which can be achieved by solving the Lagrangian dual:
\[
\text{LD} = \max_{\blambda} \text{LR}(\blambda).
\]
The OPF-DNN training solves the Lagrangian dual through a subgradient method: it
computes a sequence of multipliers
$\blambda^1,\ldots,\blambda^k,\ldots$ by solving a sequence of
trainings $\text{LR}(\blambda^0),\ldots,
\text{LR}(\blambda^{k-1}),\ldots$ and adjusting the multipliers using
constraint violations, i.e.,
\begin{align}
\bm{w}^{k+1} &= \argmin_{\bm{w}} {\cal L}[\blambda^k](\bx,\by,\hat{\cal O}[\bw^k](\bx)) \label{eq:L1} \tag{L1}\\
\bm{\lambda}^{k+1} &= \vect{\lambda^k_c + \rho\,\nu_c(\bx,\hat{\cal O}[\bw^{k+1}](\bx)) \;|\; c\in {\cal C}}. \label{eq:L2} \tag{L2}
\end{align}

\begin{algorithm}[!h]
  \caption{The Training of OPF-DNN.}
  \label{alg:learning}
  \setcounter{AlgoLine}{0}
  \SetKwInOut{Input}{input}

  \Input{${\cal D} = ({\cal X}, {\cal Y}):$ Training data\\
  		 $\alpha, \rho, u_{\lambda}$: Parameters\!\!\!\!\!\!\!\!\!\!}
  \label{line:1}
  $\lambda^1_c \gets \lambda_c^{\text{init}}, \quad \forall c \in {\cal C}$\\
  \For{epoch $k = 1, 2, \ldots$} { 
  \label{line:2}
	  \label{line:3}
	  	$\hat{\by} \gets \hat{\cal O}[\bw](\bx)$\\
  		\label{line:4}
	  	${\cal L}_o(\hat{\by}, \by) \gets \frac{1}{b}
	  	 	\sum_{\ell \in [b]} 
	  	 	{\cal L}_v(\by_\ell, \hat{\by}_\ell) + 
	  	 	{\cal L}_\theta(\by_\ell, \hat{\by}_\ell)+$\!\!\!\!\\
	  	 	\nonl
	  	 	$\hspace*{87pt}
	  	 	{\cal L}_p(\by_\ell, \hat{\by}_\ell) + 
	  	 	{\cal L}_q(\by_\ell, \hat{\by}_\ell)$\\
	  	\label{line:6}
	  	${\cal L}_c(\bx,\hat{\by}) \gets \frac{1}{b} 
	  	 	\sum_{\ell \in [b]} 
	  	 	\sum_{c \in {\cal C}} \lambda_c^k \nu_c(\bx_\ell,\hat{\by}_\ell) $\\
	  	\label{line:7}
	  	$\omega \gets \omega - \alpha \nabla_{\omega} 
	  		({\cal L}_o(\hat{\by}, \by) 
	  		+ {\cal L}_c(\bx,\hat{\by}))\!\!\!\!$
	  	        \label{line:8} \\
        $\blambda^{k+1} \gets \blambda^k$ \\
        \If{$k~\mathrm{mod}~u_{\lambda} = 0$}{                        
           \For{$c \in \mathcal{C}$}{ 
             $\lambda_c^{k+1} \gets \lambda_c^{k} + \rho \nu_c(\bx,\hat{\by})$ \label{line:9} }}
        }
\end{algorithm}

The overall training scheme is presented in Algorithm
\ref{alg:learning}.  It takes as input the training dataset ${\cal
  D}$, the optimizer step size $\alpha > 0$, the Lagrangian step size
$\rho > 0$, and the parameter $u_{\lambda}$.  The Lagrangian
multipliers are initialized in line \ref{line:1}. The training is
performed for a fixed number of epochs. The Lagrangian multipliers are
only updated after every $u_{\lambda}$ epochs which allows an accurate
approximation of (\ref{eq:L1}) to be computed before the updating
step.  After predicting the voltage and generation power quantities
(line \ref{line:4}), the objective and constraint losses are computed
(lines \ref{line:6} and \ref{line:7}). The constraint losses use the
Lagrangian multipliers $\bm{\lambda}^k$. The model weights are updated
in line \ref{line:8}.  Finally, every $u_{\lambda}$ epochs, the
Lagrangian multipliers are updated (line \ref{line:9}).

\section{Experimental Results}
\label{sec:experiments}

\subsection{Dataset Generation and ML Models}

OPF-DNN is evaluated on real-world network instances from RTE, the
French transmission operator, as well as from the NESTA
library\cite{Coffrin14Nesta} for context. Table \ref{tbl:dataset}
shows the network names and some of their characteristics. The test
cases {\tt MSR}, {\tt France EHV}, and {\tt France\_Lyon} correspond
to configurations of sub-networks of the French Transmission
system. {\tt MSR} (Marseille Sous Realtor) is the South-East region of
France, which is challenging from an operational standpoint due to
voltage issues. {\tt France\_EHV} describes the EHV level of the
French system, and {\tt France\_Lyon} is the EHV level of the French
system with a more detailed network for the Lyon region. This last
network has 3,411 buses, 4,499 lines or transformers, 3,273 loads, and
771 generators at bus-branch representation for OPF studies.

The datasets used for training and testing OPF-DNN are generated as
follows. Initially, for each load $({p}^d, {q}^d)$, lower and upper
bounds are randomly generated from two uniform distributions
\text{U}(0.8$\ell_0$, 0.9$\ell_0$) and \text{U}(1.1$\ell_0$,
1.2$\ell_0$), where $\ell_0 = ({p_0}^d, {q_0}^d)$ is the nominal
load. Then, starting from their lower bounds, the loads are uniformly
increased towards their corresponding upper bounds. Due to the
different bounds for each load, some loads have larger variations than
others. In addition, for each such snapshot, uncorrelated noise is
added to each load separately. In some cases, the network becomes too
congested and the AC-OPF (Model~\ref{model:ac_opf}) may become
infeasible, in which case the snapshot is discarded. Therefore, each
training data point consists of a load snapshot and its AC-OPF
solution. A detailed description of the generation procedure is given
in the Appendix.  The characteristics of the resulting datasets
represent realistic load profiles: the uniform load increase and the
fixed active/reactive power ratio represent the typical behavior for
aggregated demand in a given geographical region.  Randomly perturbing
each individual load in an uncorrelated fashion would produce
unrealistic load profiles. The datasets were generated by solving
AC-OPFs with the Julia package PowerModels.jl \cite{Coffrin:18} and
the nonlinear solver IPOPT \cite{wachter06on} using Intel 2.5 GHz-i7
CPUs and 16GBs of RAM.

\begin{table}[!t]
\centering
\resizebox{0.65\linewidth}{!}
  {
  \begin{tabular}{l|rrrr}
  \toprule
     \multicolumn{1}{l}{\textbf{Test Case}}   & $|{\cal N}|$ & $|{\cal E}|$ & $l$ & $g$\\
     \midrule
     \textbf{89\_pegase}   &89&210&35&12\\
     \textbf{118\_ieee}    &118&186&99&54\\
     \textbf{300\_ieee}    &300&411&201&69\\
     \textbf{MSR}  &403    &550&538&115\\
	\textbf{France\_EHV}  &1,737&2,350&1,731&290\\
   \textbf{France\_Lyon}  &3,411&4,499&3,273&771\\
   \bottomrule
  \end{tabular}
  }
  \caption{The power system networks.}
  \label{tbl:dataset} 
\end{table}

\subsection{Machine-Learning Models}

The experimental results consider three versions of OPF-DNN (Algorithm
\ref{alg:learning}) specified by three sets of values for the
hyperparameters $\blambda_{\text{init}}$ and $\rho$:
\begin{itemize}[wide, labelwidth=!, labelindent=0pt]
\item $\mathcal{M}_B$ $(\blambda^{\text{init}}\!=\!0,\rho\!=\!0)$: the
  base model that ignores the OPF constraints.
  
\item $\mathcal{M}_C$ $(\blambda^{\text{init}}\!=\!1,\rho\!=\!0)$: the
  base model with constraint violations in the objective but no
  updates of Lagrangian multipliers.
\item $\mathcal{M}_C^D$ $(\blambda^{\text{init}}\!=\!0,
  \rho)$: the proposed OPF-DNN. 
\end{itemize}
\noindent
The ML models were implemented in Python 3.0 using PyTorch
\cite{paszke:17}. The training was performed using NVidia Tesla V100
GPUs with $16$ GB of memory, and the Adam optimizer with learning rate
$lr = 10^{-4}$ and a maximum of $50000$ epochs.  The Lagrangian step
size $\rho$ is set to a value in the interval $[10^{-4}, 10^{-2}]$ and
the multipliers are updated every $1000-5000$ epochs. These
hyperparameters were tuned for each benchmark. The evaluation study
uses a $80/20$ train-test split and reports results only on the test
set. For all test cases, the dataset size $N$ was set to $2000$.

\subsection{Model Comparison}

\begin{figure*}[!t]
\centering
\footnotesize{MSR}\\
\includegraphics[width=0.9\linewidth]{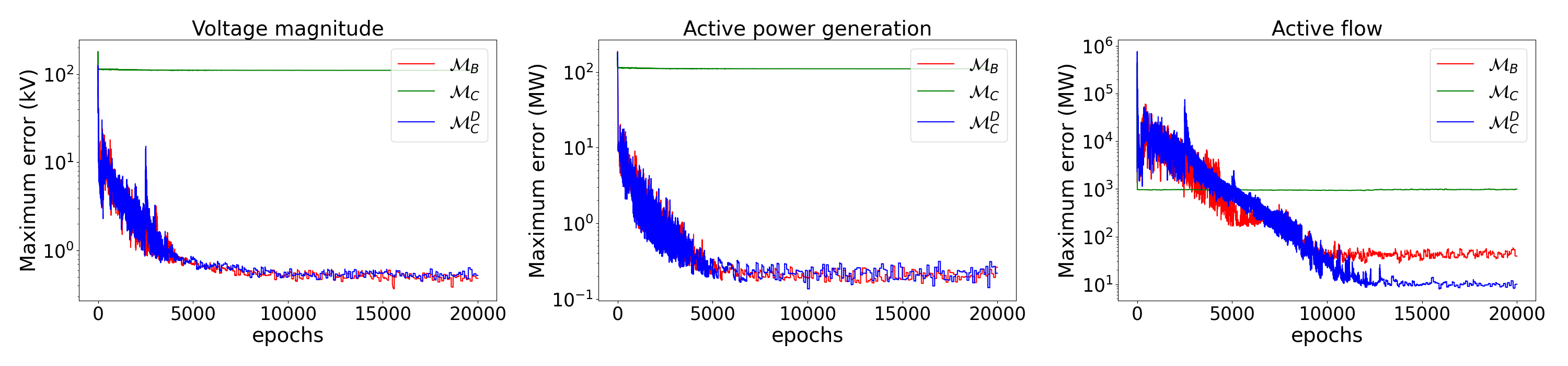}\\
\footnotesize{France\_EHV}\\
\includegraphics[width=0.9\linewidth]{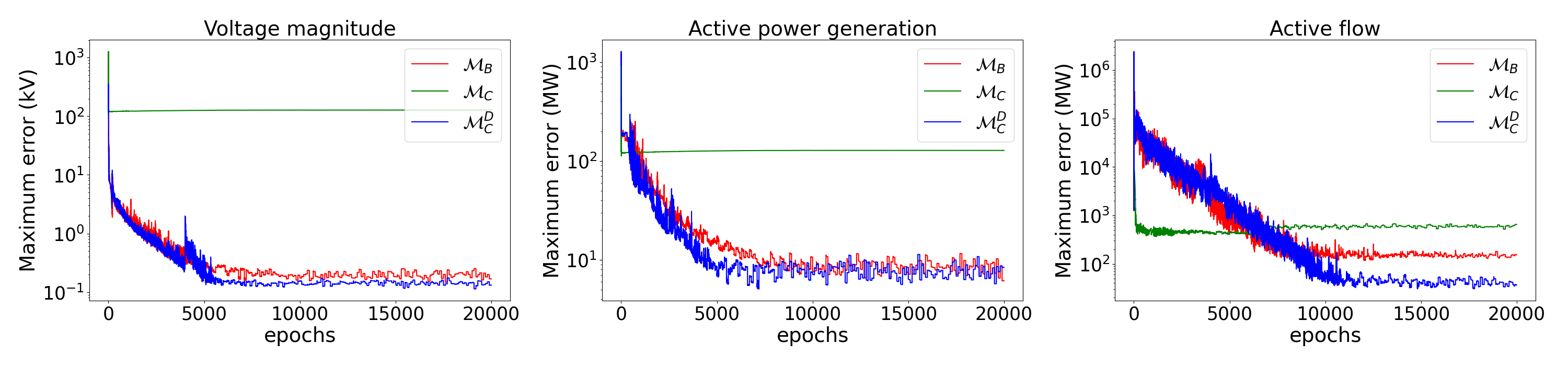}\\
\footnotesize{France\_Lyon}\\
\includegraphics[width=0.9\linewidth]{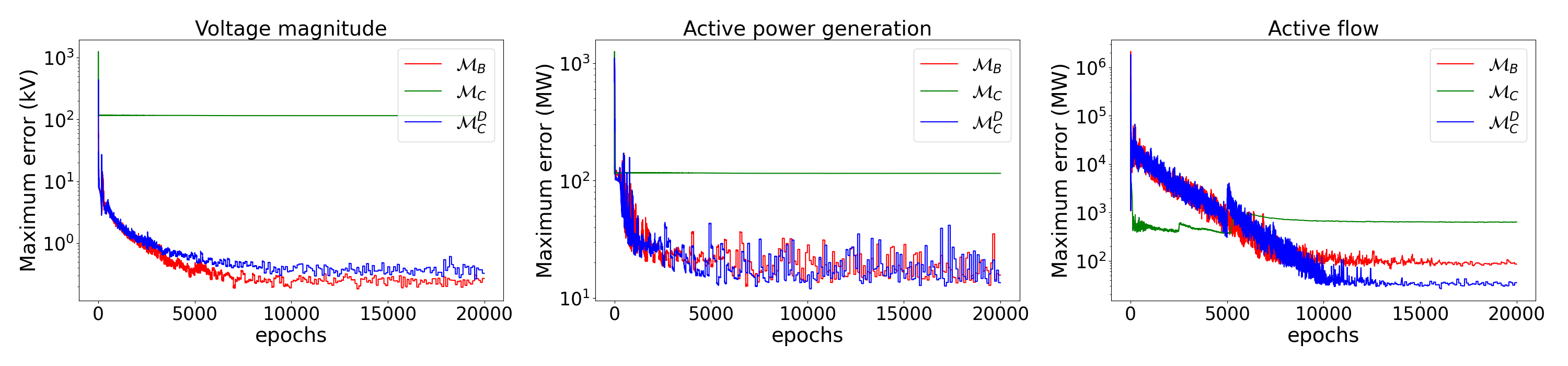}\\
\caption{Convergence analysis for $\mathcal{M}_B, \mathcal{M}_{C}, \mathcal{M}_{C}^{D}$ on the French Transmission System topologies.}
\label{fig:convergence}
\end{figure*}

\begin{figure*}[!t]
\centering
\footnotesize{MSR}\\
\includegraphics[width=0.9\linewidth]{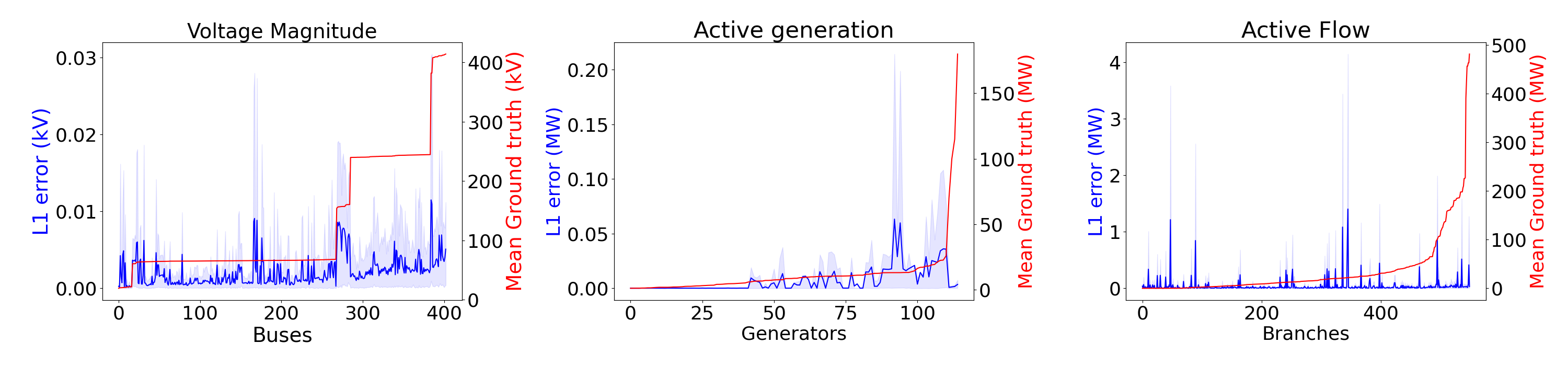}\\
\footnotesize{France\_EHV}\\
\includegraphics[width=0.9\linewidth]{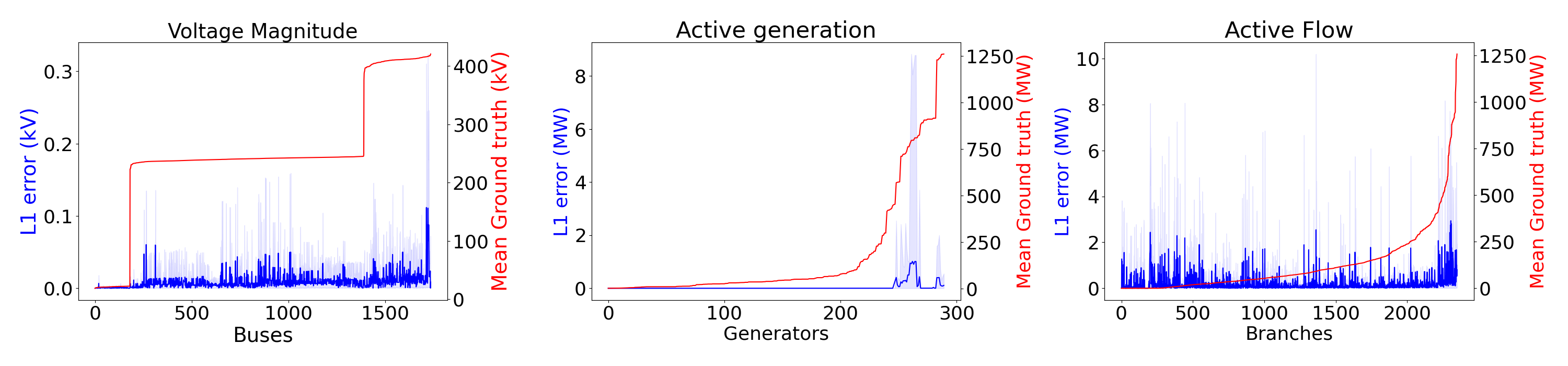}\\
\footnotesize{France\_Lyon}\\
\includegraphics[width=0.9\linewidth]{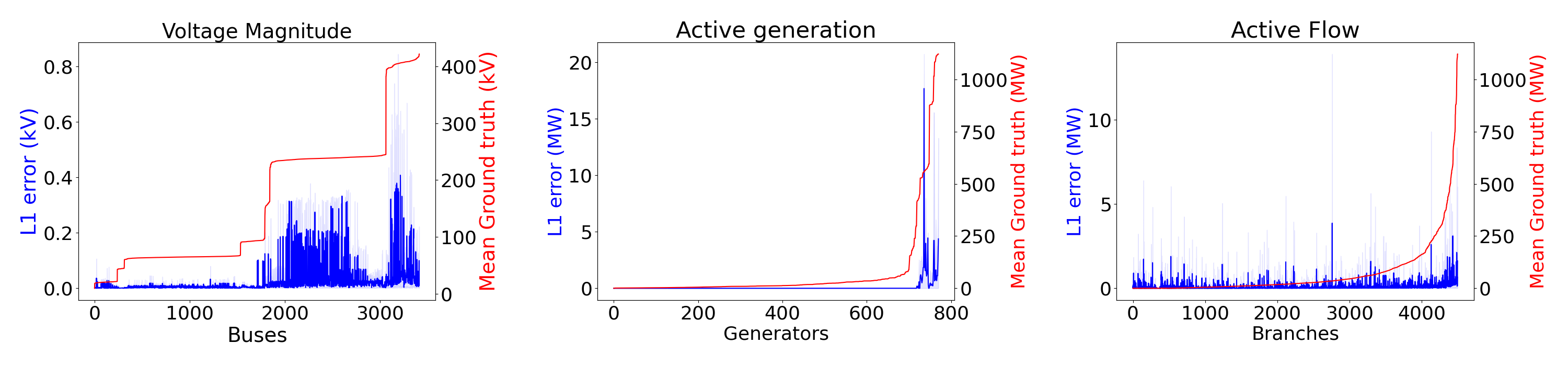}
\caption{Prediction Errors over Buses, Generators, and Branches. The Ground Truth and Error Graphs are in Different Scales.}
\label{fig:prediction_errorsls}
\end{figure*}

\noindent
Figure \ref{fig:convergence} displays, for each learning algorithm,
the convergence of the training procedure for voltage magnitudes,
active power generations, and active flows on the topologies of the
French transmission system. It reports the maximum average error for
buses, generators, and branches over all training instances. The
figure highlights that $\mathcal{M}_C$ fails to converge to a
satisfying point for the three predictions. This behavior is due to
the non-convex part of the loss function stemming from constraints
\eqref{con:5a} and \eqref{con:5b}. The error for each such constraints
is extremely large ($10^3$ in p.u., on average), leading to exploding
gradients and an inefficient training procedure. In contrast,
$\mathcal{M}_B$ and $\mathcal{M}_C^{D}$ exhibit similar convergence
profiles for voltage and active power values.  However,
$\mathcal{M}_C^{D}$ enjoys stronger convergence behaviors for branch
flows, which are indirectly predicted through Ohm's law. The more
accurate prediction of branch flows is crucial to satisfy the AC-OPF
constraints.

\subsection{Prediction Errors}

The predictive accuracy of OPF-DNN is assessed by comparing the
directly predicted values $(\hat{\bm{v}}, \hat{\bm{\theta}},
\hat{\bm{p}}^g,\hat{\bm{q}}^g)$, as well as the indirectly predicted
value $\hat{\bm{p}}^f$, to their respective ground truth values
$(\bm{v}, \bm{\theta}, \bm{p}^g, \bm{q}^g, \bm{p}^f)$. Figure
\ref{fig:prediction_errorsls} depicts the absolute errors for voltage
magnitude $\bm{v}$, active generation $\bm{p}$, and active flow
$\bm{p^f}$ for the French benchmarks. To communicate the results more
clearly, the buses, generators, and branches are \emph{sorted in
  ascending orders} of their mean ground-truth values. The figure
reports the mean $L_1$-error over the test cases as well as intervals
that capture $95 \%$ of the test cases errors. For the voltage
magnitudes, the mean error is under $0.1\%$ of the nominal value. For
the active power, the mean error is extremely low and, in almost all
cases, well below $1\%$ of the nominal value. The intervals indicate
that there are some cases with larger errors. These cases are
instances with large loads, which induce significant changes in
solution values.  Adding more instances in these operating regions
would improve results. Finally, OPF-DNN yields extremely low errors
(well under $1\%$) for the indirectly predicted active flows for lines
carrying large flows (typically $90\%$ of the lines).  For lines with
small flows, although the absolute error is still very low, it is
larger in percentage, as the training minimizes the error in absolute
scale (but not in relative scale). Branch errors can be strongly
correlated with their impedance values, as impedance values directly
affect the power flows (indirect) predictions. Table
\ref{tbl:predictionerrorall} reports comprehensive results on the
absolute errors. For each quantity $\bx$, it reports the value $
\frac{1}{m N}\lvert \lvert \bx - \hat{\bx} \rvert \rvert_1$, where $m$
is the number of buses, generators, or branches.  The results are
consistent across all the benchmarks and demonstrate the high accuracy
of the OPF-DNN predictions.

\begin{table}[!ht]
\centering
\small
\resizebox{\linewidth}{!}
  {
  \begin{tabular}{l|rrrrrr}
  \toprule
     \multicolumn{1}{l}{\textbf{Test Case}}   & $\hat{\bm{v}}$ &  $\hat{\bm{\theta}}$ & $\hat{\bm{p}}^g$ & $\hat{\bm{q}}^g$ & $\hat{\bm{p}}^f$ & $\hat{\bm{q}}^f$\\
     \midrule
     \textbf{89\_pegase}  &0.025&0.005&0.585&0.425&0.152&0.353\\
     \textbf{118\_ieee}  &0.009&0.004&0.013&0.016&0.035&0.059\\
     \textbf{300\_ieee}  &0.006&0.020&0.183&0.068&0.152&0.101\\
     \textbf{MSR}  &0.002&0.001&0.007&0.005&0.039&0.066\\
	\textbf{France\_EHV}  &0.012&0.008&0.032&0.104&0.146&0.099\\
   \textbf{France\_Lyon}  &0.025&0.005&0.098&0.093&0.087&0.153\\
   \bottomrule
  \end{tabular}}
  \caption{Mean prediction errors in kV, degrees, MW, MVA for $\hat{\bm{v}}, \hat{\bm{\theta}}, (\hat{\bm{p}}^g, \hat{\bm{p}}^f),(\hat{\bm{q}}^g, \hat{\bm{q}}^f$), respectively.}
  \label{tbl:predictionerrorall} 
\end{table}

\subsection{Feasibility Errors}

\begin{figure}[!th]
\includegraphics[width=\columnwidth]{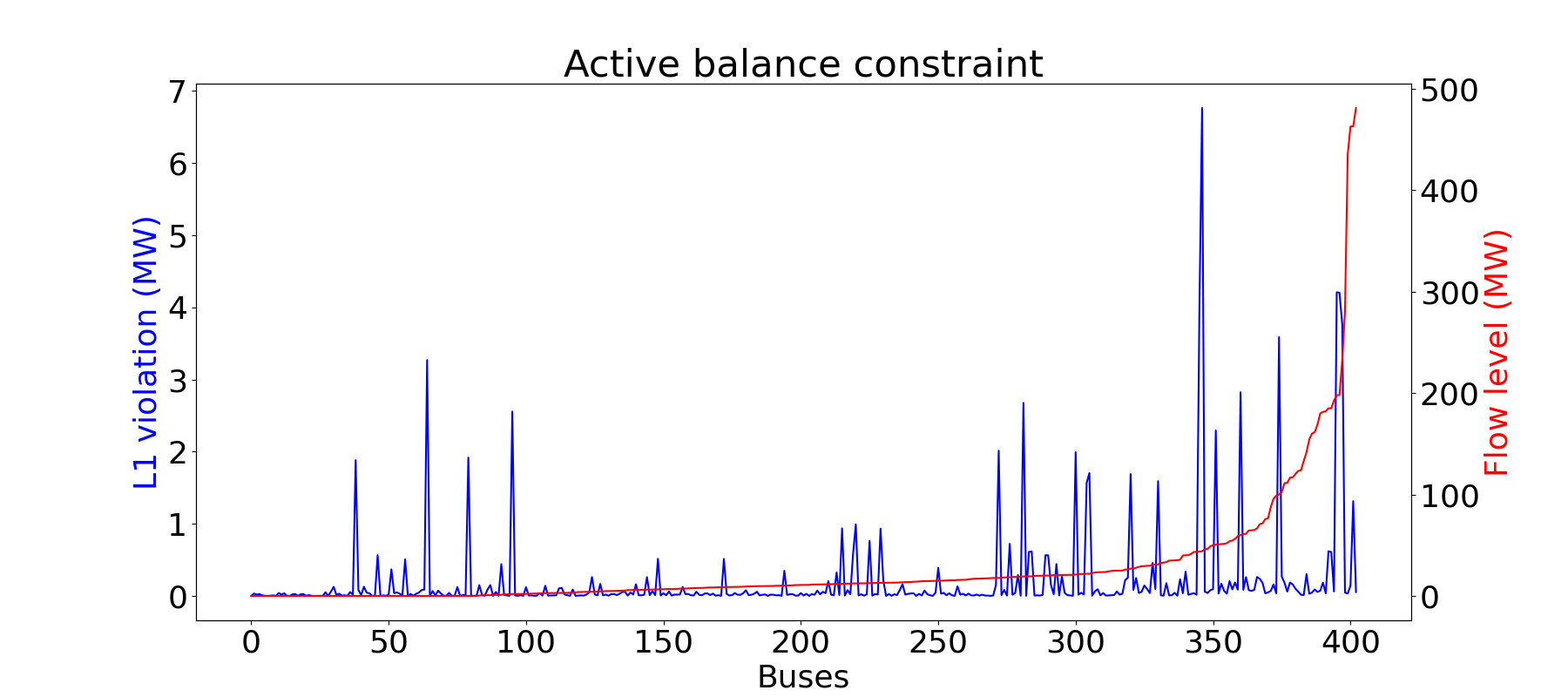}
\caption{\label{fig:activemsr} Mean absolute violations of the
  active power balance constraint \eqref{con:6a} for the MSR test case across all buses. The ground truth and the violations are in different scales.}
\end{figure}

It is important to study how the predictions $(\hat{\bm{v}},
\hat{\bm{\theta}}, \hat{\bm{p}}^g,\hat{\bm{q}}^g, \hat{\bm{p}}^f)$
sastify the OPF constraints. There are no violations due to
constraints \eqref{con:5a} and \eqref{con:5b} since these equations
(Ohm's law) are used to compute the predicted flow values
$(\hat{\bm{p}}^f, \hat{\bm{q}}^f)$. Table \ref{tbl:feasibilityall}
reports results on the satisfaction of constraints \eqref{con:2a},
\eqref{con:3a}, \eqref{con:3b}, and \eqref{con:4} in percentage.
Table \ref{tbl:feasibilitybound2} complements these results and
reports the mean violations for the violated constraints in
Table~\ref{tbl:feasibilityall}.  Power balance constraints
\eqref{con:6a} and \eqref{con:6b} are also included. The mean
violations are small, indicating the strong accuracy of the OPF-DNN
predictions. Moreover, the violation pattern over the buses for
constraints \eqref{con:6a} displays the same behavior as that observed
for the errors on the branch flows: the errors tend to be very small
for buses adjacent to branches carrying large flows.  Figure
\ref{fig:activemsr} visualizes this information for a representative
case study: it reports the mean absolute violations of constraints
\eqref{con:6a} for the MSR test case. Observe how small the constraint
violations are in absolute terms.

\begin{table}[!t]
\centering
\small
\resizebox{0.75\linewidth}{!}
  {
  \begin{tabular}{l|rrrrrr}
  \toprule
     \multicolumn{1}{l}{\textbf{Test Case}}   & \eqref{con:2a} & \eqref{con:3a} & \eqref{con:3b} & \eqref{con:4} \\
     \midrule
     \textbf{89\_pegase}  &99.9&99.8&98.8&99.9\\
     \textbf{118\_ieee}  &99.9&99.7&99.8&100.0\\
     \textbf{300\_ieee}  &98.7&98.3&98.9&99.7\\
     \textbf{MSR}  &99.8&99.6&99.9&100.0\\
	\textbf{France\_EHV}  &99.7&99.2&99.7&100.0\\
   \textbf{France\_Lyon}  &99.8&99.8&99.9&100.0\\
   \bottomrule
  \end{tabular}}
  \caption{Percentage of bound constraints satisfied.}
  \label{tbl:feasibilityall} 
\end{table}

\begin{table}[!th]
\centering
\small
\resizebox{0.9\linewidth}{!}
  {
  \begin{tabular}{l|rrrrrr}
  \toprule
     \multicolumn{1}{l}{\textbf{Test Case}}   & \eqref{con:2a} & \eqref{con:3a} & \eqref{con:3b} & \eqref{con:4} & \eqref{con:6a} & \eqref{con:6b}\\
     \midrule
     \textbf{89\_pegase}  &$<$0.01&4.52&0.45&1.39&0.25&1.01\\
     \textbf{118\_ieee}  &$<$0.01&1.83&0.18&-&0.04&0.10\\
     \textbf{300\_ieee}  &$<$0.01&0.43&0.20&0.21&0.16&0.17\\
     \textbf{MSR}  &$<$0.01&0.62&0.12&-&0.09&0.15\\
	 \textbf{France\_EHV}  &0.01&2.94&0.79&-&0.24&0.14\\
     \textbf{France\_Lyon}  &$<$0.01&1.30&0.50&-&0.16&0.16\\
   \bottomrule
  \end{tabular}}
  \caption{Mean violation for violated AC-OPF constraints. The
    violation is expressed in kV, MW and MVA for constraints
    \ref{con:2a}, (\ref{con:3a}, \ref{con:6a}) and (\ref{con:3b},
    \ref{con:4}, \ref{con:6b}), respectively.}
  \label{tbl:feasibilitybound2} 
\end{table}

\subsection{Objective Values}

Table \ref{tbl:objectiveerrors} compares the objective values 
from the AC-OPF optimizations with those obtained by the OPF-DNN predictions. 
In addition, the table also compares the costs obtained from the 
load flow model (\ref{model:load_flow}) seeded with 
the OPF-DNN predictions. In
particular, Table \ref{tbl:objectiveerrors} reports the average
relative objective error
\(
\frac{1}{N_{\text{test}}}\sum_{i=1}^{N_{\text{test}}} |1 - \frac{\hat{c_i}}{c_i}| \times 100,
\)
where $c_i$ is the AC-OPF objective value and $\hat{c}_i$ is either
the objective value produced by OPF-DNN or by the load flow
solution.
The results demonstrate that the OPF-DNN produces objective
values closely approximate the AC-OPF costs. 
The closest feasible solutions obtained by solving load flow 
are also extremely accurate.

\begin{table}[!th]
\centering
\small
\resizebox{0.7\linewidth}{!}
  {
  \begin{tabular}{l|rr}
  \toprule
     \multicolumn{1}{l}{\textbf{Test Case}}  & OPF-DNN & Load Flow\\
     \midrule
     \textbf{89\_pegase}  &$25 \cdot 10^{-3}$&$0.13 \cdot 10^{-3}$\\
     \textbf{118\_ieee}  &$16\cdot 10^{-3}$&$0.18 \cdot 10^{-3}$\\
     \textbf{300\_ieee}  &$10\cdot 10^{-3}$&$0.42 \cdot 10^{-3}$\\
     \textbf{MSR}  &$6\cdot 10^{-3}$&$0.21\cdot 10^{-3}$\\
	\textbf{France\_EHV}  &$5\cdot 10^{-3}$&$0.16 \cdot 10^{-3}$\\
   \textbf{France\_Lyon}  &$25\cdot 10^{-3}$&$2.75\cdot 10^{-3}$\\
   \bottomrule
  \end{tabular}}
  \caption{Differences (in \%) between the OPF, the OPF-DNN objective values and the associated load flow objective.}
  \label{tbl:objectiveerrors} 
\end{table}

\subsection{Time and Memory}

Finally, table \ref{tbl:implementation} reports the average computation times and the GPU memory utilization for the OPF-DNN training across all benchmarks. In addition, it reports the time needed to produce a prediction for a given instance. 
To give the proper context, the table also reports the average time to solve the AC-OPFs. For the case studies, the training times are reasonable and the prediction times are up to 4 orders of magnitude faster than the AC-OPF runtimes. \emph{OPF-DNN, thus, provides an appealing tradeoff between accuracy and efficiency}.
\begin{table}[!t]
\centering
\small
\resizebox{\linewidth}{!}
  {
  \begin{tabular}{l|ccc|cc}
  \toprule
        & Train time &  Predict time & Train Mem. & AC-OPF \\
    \textbf{Test Case}   & (min) &  (sec) & (GB) & (sec) \\
     \midrule
     \textbf{89\_pegase}  &48&0.0013&1.1&0.2\\
     \textbf{118\_ieee}  &51&0.0013&1.1&0.2\\
     \textbf{300\_ieee}  &54&0.0014&1.2&1.9\\
     \textbf{MSR}  &59&0.0016&1.5&2.2\\
	 \textbf{France\_EHV}  &142&0.0016&4.6&4.2\\
     \textbf{France\_Lyon}  &444&0.0020&13.9&47.9\\
   \bottomrule
  \end{tabular}}
  \caption{Training and inference computational costs.}
  \label{tbl:implementation} 
\end{table}

\section{Conclusions}

The paper proposed OPF-DNN, a deep learning approach to produce
high-fidelity approximations of large-scale optimal power flows in
milliseconds.  The approach combines deep learning and Lagrangian
duality to model the physical and engineering constraints.
Computational experiments on real-life case studies based on the
French transmission system (up to 3,400 buses and 4,500 lines)
demonstrate the potential of the approach. These results open a new
avenue in approximating the AC-OPF, a key building block in many power
system applications, including expansion planning and security
assessment studies. Current work aims at scaling the approach to even
larger systems with over $10^4$ buses, which raise challenges in GPU
memory and training time.


\newpage 
\appendix

Algorithm \ref{alg:dataset} presents the details of the data generation procedure.

\begin{algorithm}[!hbt]
  \caption{Dataset generation}
  \label{alg:dataset}  
  \setcounter{AlgoLine}{0}
  \SetKwInOut{Input}{input}
  \Input{$N$: The size of the dataset, \\
  $\bm{\ell_0} = (\bm{p_0}^d, \bm{q_0}^d)$: The nominal load}
  \For{$\ell$ in $[n]$}
  { 
     $\bm{lb}[\ell] \gets U[0.8\bm{\ell_0}(\ell), 0.9\bm{\ell_0}(\ell)]$, 
     
     $\bm{ub}[\ell] \gets U[1.1\bm{\ell_0}(\ell), 1.2\bm{\ell_0}(\ell)]$
  }
  \For{$c = 0, \frac{1}{N}, \frac{2}{N}, ..., 1$} { 
  $\bm{load} \gets (1-c) \ \bm{lb} + c \ \bm{ub}$. \\
	\For{$\ell$ in $[n]$}
  { 
    $\xi \gets U[(1 - \frac{\bm{ub}[\ell] - \bm{lb}[\ell]}{100N}), (1 + \frac{\bm{ub}[\ell] - \bm{lb}[\ell]}{100N})]$
    
     $\bm{load}[\ell] \gets \xi \cdot \bm{load}[\ell]$
  }
    result = AC-OPF ($\bm{load}$) \\
    \If{AC-OPF is feasible}{ $\mathcal{D}$ = Union ($\mathcal{D}$, (result, $\bm{load}$))}
  }
\end{algorithm}

\end{document}